\documentclass[aps,prl, reprint,superscriptaddress, preprintnumbers]{revtex4-1}
\usepackage{blindtext}
\usepackage[T1]{fontenc}
\usepackage[english]{babel}
\usepackage{amsmath}
\pdfoutput=1 

\def\be{\begin{equation}}
\def\ee{\end{equation}}
\def\bea{\begin{eqnarray}}
\def\eea{\end{eqnarray}}
\def\pd{\partial}
\def\a{\alpha}
\def\b{\beta}

\def\d{\delta}
\def\m{\mu}
\def\n{\nu}

\def\l{\lambda}

\def\r{\rho}

\def\s{\sigma}
\def\e{\epsilon}

\def\bma{\begin{pmatrix}}
\def\ema{\end{pmatrix}}

\def\bi{\begin{itemize}}
\def\ei{\end{itemize}}

\begin{document}
	\preprint{IFT-UAM/CSIC-15-064}
	\preprint{FTUAM-15-18}
	
\title{First Order formulation of Unimodular gravity}
\author{E. Álvarez}
\email{enrique.alvarez@uam.es}
\affiliation{Instituto de Física Teórica UAM/CSIC, C/ Nicolas Cabrera, 13-15, C.U. Cantoblanco, 28049 Madrid, Spain}
\affiliation{Departamento de Física Teórica, Universidad Autónoma de Madrid, 20849 Madrid, Spain}
\author{S. González-Martín}
\email{sergio.gonzalez.martin@csic.es}
\affiliation{Instituto de Física Teórica UAM/CSIC, C/ Nicolas Cabrera, 13-15, C.U. Cantoblanco, 28049 Madrid, Spain}
\affiliation{Departamento de Física Teórica, Universidad Autónoma de Madrid, 20849 Madrid, Spain}

\begin{abstract}
First order lagrangians for the  Weyl invariant formulation of Unimodular Gravity are proposed. Several alternatives are examined; in some of them first and second order are equivalent in a certain gauge only.
\end{abstract}
\maketitle
\section{Introduction}

Unimodular Gravity (cf. \cite{AlvarezF} and references therein) is undoubtedly a quite interesting theory. It has been recently studied in its second order Weyl invariant \cite{AlvarezHV,AlvarezGMHVM} formulation.
The aim of the present paper is to  consider UG in first order formulation. The first order Palatini action principle \cite{Palatini} which is classically equivalent to the corresponding one for General Relativity contains two independent fields, namely the frame field defined by the tetrad  $e_a^\m(x)$ and the spin connection $\omega^{ab}_\m(x)$ and reads
 \be
 S=-{1\over 2\kappa^2}\int d^4 x ~e~e_a^\m ~e_b^\n~ R_{\m\n}^{ab}\left[\omega\right]
\ee
where
\be
e\equiv \left|\text{det}~e^a_\m\right|
\ee
\be
R_{\m\n}^{ab}\left[\omega\right]\equiv \pd_\m \omega^{ab}\;_\n-\pd_\n\omega^{ab}\;_\m+\omega^a\,_{c\m}\omega^{cb}\,_{\n}-\omega^a\,_{c\n}\omega^{cb}\,_{\m}
\ee
In terms of differential forms
\be
R^{ab}\equiv d\omega^{ab}+\omega^a\,_c\wedge \omega^c\,_b
\ee
and the action itself can be written \cite{note} as
\be
S=-{1\over 8\kappa^2}\int \e_{abcd}~e^a\wedge e^b\wedge R^{cd}\equiv -{1\over 8\kappa^2}\int *\left(~e^a\wedge e^b\right)\wedge R^{cd}
\ee
In this paper we are going to consider only pure gravity with an eye on extensions to Unimodular Supergravity \cite{Nishino} which is presumably easier to study in its first order form. 

\par
When $e\neq 0$  then
the equation of motion (EM) for the spin connection forces the torsion
\be
T^a\equiv d e^a+\omega^a\,_b\wedge e^b
\ee
to vanish, and this forces the said connection to be identified with the Ricci rotation coefficients. The equation of motion for the frame field then gives the Ricci flatness condition. It is possible to get solutions even when $e=0$ \cite{Tseytlin}, which are quite interesting although not for the purposes of the present paper. Incidentally, this shows that the first order action principle is slightly more general than the second order one.
\par
In order to get a first order action in the unimodular case   \cite{Blas} we simply write
\be
S=-{1\over 2\kappa^2}\int d^4 x ~\hat{e}_a^\m ~\hat{e}_b^\n~ R_{\m\n}^{ab}\left[\omega\right]
\ee
where the frame field $\hat{e}_a^\m$ is assumed to have unit determinant as and ordinary matrix. The action can be written in terms of an arbitrary frame field 
\be
e_a^\m(x)\equiv e^{- 1/4} \hat{e}_a^\m
\ee
as
\bea
S&&=-{1\over 2\kappa^2}\int d^4 x ~e^{1/2}~e_a^\m ~e_b^\n~ R_{\m\n}^{ab}[\omega]=\nonumber\\&&=-{1\over 8\kappa^2}\int  ~e^{- 1/2}~e^a\wedge e^b\wedge~ R^{cd}\e_{abcd}
\eea
This action, besides being Lorentz and Diff invariant, is also  Weyl invariant under

\be
e^a_\m\rightarrow \Omega(x) e^a_\m
\ee
Our first task is to carefully derive the EM for this theory.
\section{Weyl invariant lagrangian with the spin connection as a Weyl singlet}
 First of all, let us consider the EM for the frame field. Assuming $e\neq 0$, as it is done throughout this work, it reads
\be
\d S=-\int d^4 x~e^{1/2}~\left(-{1\over 2}~e^d_\a e_a^\l e_b^\s R^{ab}_{\l\s}+ e_b^\n R^{db}_{\a\n}+ e_a^\m R^{ad}_{\m\a}\right)\d e^\a_d
\ee
The origin of the minus sign is the identity
\be
e_d^\a \d e^d_\a=-e^d_\a \d e^\a_d
\ee
Multiplying by the frame field $e_d^\a$ we get an identity; that is the EM are traceless in the sense that
\be
e_a^\l~{\d S\over \d e_a^\l}\equiv 0
\ee
The  nontrivial piece tells us that
\be
R^{d b}_{\a \n} ~e^\n_b-{1\over 4}~\left(e_a^\l e_b^\s R^{ab}_{\l\s}\right)~e^d_\a=0
\ee
This EM coincides with the one obtained in second order formalism; the only thing is that the spin connection is not yet determined.

 The  variation of the connection gives 
\be
\d S=\int d^4 x~ e^{1/ 2}~ e_a^\m e_b^\n \d R^{ab}_{\m\n}
\ee
where
\be
\d R^{ab}=D \d\omega^{ab}
\ee
and $D$ represents the Lorentz covariant derivative.
\par
The variations $\d\omega^{ab}$ are Lorentz tensors so that the whole expression can be integrated by parts
\be
\d S=-\int \e_{abcd}~D\left(e^{-{1/ 2}} e^a\wedge e^b\right)\d \omega^{cd}
\ee
The Lorentz covariant derivative of the frame itself is nothing else than the two-form {\em torsion}, which is a Lorentz vector
\be
T^a\equiv D e^a \equiv d e^a+ \omega^a_b \wedge e^b
\ee
 The torsion two-form transforms nonlinearly under Weyl on the assumption that the spin connection remains inert.
 \be
 T^a\rightarrow \Omega T^a+d\Omega\wedge e^a
 \ee
 The result is actually quite simple. In terms of the unimodular frame the variation is {\em exactly} as in the Palatini case, to that
 \be
 \hat{T}^a=0
 \ee
 Weyl transforming with
 \be
 \Omega\equiv e^{1/ 4}
 \ee
 then yields the non-vanishing torsion when arbitrary frames are considered.
 \par
 Let us do the explicit calculation to check our result. Taking into account that
\be
 d e= e ~e_a^\m ~d e^a_\m
\ee
it follows
\be
\d S=\int \e_{abcd}e^{-{1/ 2}}~\left(-{1\over 2}{d e\over e}\wedge e^a\wedge e^b + T^a\wedge e^b-e^a\wedge T^b\right)\d \omega^{cd}
\ee
Disentangling the EM
\bea
\e_{abcd}&&\bigg\{T^c_{\m\n} e^d_\l+T^c_{\l\m}e^d_\n + T^c_{\n\l} e^d_\m-{1\over 2} e_k^\s\left(\pd_\m e^k_\s e^c_\n e^d_\l+\right.\nonumber\\ &&\left.\pd_\l e^k_\s e^c_\m e^d_\n +\pd_\n e^k_\s e^c_\l e^d_\m\right)\bigg\}=0
\eea
Multiplying by $ e_g^\l e_e^\n e_f^\m$
\bea
\e_{abcd}&&\bigg\{T^c_{f e} \d^d_g+T^c_{g f} \d^d_e + T^c_{e g } \d^d_f-{1\over 2} e_k^\s\left(\pd_f e^k_\s \d^c_e \d^d_g+\pd_g e^k_\s \d^c_f \d^d_e \right.\nonumber\\&&\left.+\pd_e e^k_\s \d^c_g \d^d_f\right)\bigg\}=\e_{abcg} T^c_{fe}+\e_{abce} T^c_{gf} +\e_{abcf} T^c _{eg}-\nonumber\\
&&-{1\over 2} e_k^\s \left(\pd_f e^k_\s \e_{abeg}+\pd_g e^k_\s \e_{abfe}+\pd_e e^k_\s \e_{abgf}\right)=0
\eea
Multiplying by $\e^{abmn}$ yields
\bea
&&T^c_{fe}\left(\d_c^m \d_g^n-\d_c^n\d_g^m\right)+T^c_{gf}\left(\d^m_c\d^n_e-\d^m_e\d^n_c\right)+\nonumber\\&&+T^c_{eg}\left(\d^m_c\d^n_f-\d^m_f\d^n_c\right)-{1\over 2}e_k^\s\times\bigg\{\pd_f e^k_\s\left(\d^m_e\d^n_g-\d^m_g\d^n_e\right)+\nonumber\\
&&+\pd_g e^k_\s\left(\d^m_f \d^n_e- \d^m_e \d^n_f\right)+\pd_e e^k_\s\left(\d^m_g \d^n_f- \d^m_f \d^n_g\right)\bigg\}=0
\eea
Finally, multiplying by $\d_n^g$
\bea
&& 4 T^m_{f e}-T^m_{f e}+T^m_{ef}-T^i_{i f}\d^m_e + T^m_{ef}-T^i_{e i}\d^m_f-\nonumber\\&&- e_k^\s\left(\pd_f e^k_\s\d^m_e-\pd_e e^k_\s\d^m_f\right)=0
\eea
Let us dub
\be
T_e\equiv T^i_{i e}=-T^i_{e i}
\ee
Taking the trace of the last equation 
\be
T_e=-\dfrac{3}{4} e_k^\s \pd_e e^k_\s
\ee
Then
\bea
T^m_{f e}&&={3\over 4}\left( e_k^\s \pd_e e^k_\s \d^m_f-e_k^\s \pd_f e^k_\s \d^m_e\right)+ \left(e_k^\s \pd_f e^k_\s \d^m_f-e_k^\s \pd_e e^k_\s \d^m_e\right)= \nonumber\\
&&=-{1\over 4}\left( e_k^\s \pd_e e^k_\s \d^m_f-e_k^\s \pd_f e^k_\s \d^m_e\right)=\nonumber\\&&=-{1\over 4}\left( \d^u_e\d^m_f-\d^u_f\d^m_e\right)e^{-1}\pd_u e
\eea

It is easy to check that the trace $T_e$ is consistent with it.
It has been already pointed out that under a Weyl transformation, the on-shell spacetime torsion is not invariant, but rather,
\bea
\overline{T}^\a_{\l\b}&&=T^\a_{\l\b}-\Omega^{-1}\left(\d^\a_\l \pd_\b\Omega  -\d^\a_\b \pd_\l \Omega \right)\nonumber\\&&=T^\a_{\l\b}-\left(\d^\a_\l \d^\s_\b -\d^\a_\b\d^\s_\l \right)\Omega^{-1}\pd_\s \Omega 
\eea
Clearly the torsion vanishes  in the unimodular gauge
\be
e=1
\ee
It is however somewhat disturbing that it does not vanish in a general gauge.

\section{Weyl invariant lagrangian with the spin connection Weyl non-singlet}
It is possible, and maybe more natural, to impose that after a Weyl transformation the spin connection remains torsion free. This imposes the transformation law
\be\label{spin}
d\left(\Omega e^a\right)+\tilde{\omega}^a\,_b\wedge \left(\Omega e^b\right)=0
\ee
This leads to a specific Weyl transformation law for the spin connection namely
\be\label{variant}
\tilde{\omega}_{abc}=\Omega^{-1}\left(\omega_{abc}+{1\over 2}\left(\pd_b\text{log}~\Omega~\eta_{ac}-\pd_a~\text{log}~\Omega ~\eta_{bc}\right)\right)
\ee
There is then a  Weyl invariant unimodular connection  given by
\be
\hat{\omega}_{abc}=e^{1/ n}\left(\omega_{abc}+{1\over 2 n}\left(\eta_{bc}~\pd_a\text{log}~e-\eta_{ac}~ \pd_b\text{log}~e\right)\right)
\ee
It is easy to check that this construct is indeed Weyl invariant, provided the spin connection does transform as in \eqref{variant}
\be
\tilde{\hat{\omega}}_{abc}=\hat{\omega}_{abc}
\ee
Let us now consider the first order action given by

\be
S=-{1\over 2\kappa^2}\int d^n x ~\hat{e}_a^\m ~\hat{e}_b^\n~ R_{\m\n}^{ab}\left[\hat{\omega}\right]
\ee
Now the same argument as before shows that the torsion vanishes. Namely, perform the variations with respect to the Weyl invariant spin connection, $\d \hat{\omega}$ (they are as arbitrary as $\d \omega$). Then we learn as before that the torsion expressed in terms of the unitary frame vanishes
\be
\hat{T}^a=0
\ee
But now the torsion is Weyl invariant so that the torsion also vanishes in a general gauge. 
\par
There is now however no reason for  the graviton EM to be traceless, because the Weyl invariant spin connection depends explicitly on the variable $e$.
The resulting EM is

\be
R^{d b}_{\a \n} ~e^\n_b-{1\over 4}~\left(e_a^\l e_b^\s R^{ab}_{\l\s}\right)~e^d_\a+\dfrac{45}{32}\dfrac{\partial_\r e\pd^\r e}{e^2} e^d_\a-\dfrac{15}{8}\dfrac{\partial_\m\pd^\m e}{e} e^d_\a=0
\ee
which reduces to the second order unimodular one in the unimodular gauge $e=1$. This particular lagrangian is somewhat unnatural in that it depends not only on $\hat{e}^a$ but also on $e$.
\section{Weyl variant lagrangian}
Nothing prevents us however to write a lagrangian like
\be
S=-{1\over 2\kappa^2}\int d^n x ~\hat{e}_a^\m ~\hat{e}_b^\n~ R_{\m\n}^{ab}\left[\omega\right]
\ee
where the spin connection is gauge variant as in \eqref{variant}. In that way we recover the traceless EM for the graviton (because the lagrangian depends on $\hat{e}^a$ only), 
and the vanishing of the torsion is a Weyl gauge invariant statement. The action itself is not, however, Weyl invariant. This fact should not be contemplated as
 a drawback; after all, Weyl invariance in our approach is  simply an artifact in order to construct a unimodular frame field out of a general one.

\section{Conclusions}
Several alternatives for first order lagrangians for Unimodular Gravity are discussed. The first one is  Weyl invariant with spin connection behaving as a Weyl singlet. 
It  does imply a nonvanishing value for the torsion in a general Weyl gauge. This is at variance what is known from the second order approach. They are certainly not fully equivalent in the present formulation.
\par
It is possible to postulate a transformation law for the spin connection in such a way that the torsion field is Weyl invariant. The corresponding Weyl invariant lagrangian produces traceful 
graviton EM. 

It is however easy to build a non Weyl-invariant first order lagrangian  in such a way that the corresponding EM are  equivalent to the  Weyl invariant second order one.

\section{Acknowledgments}
 This work has been partially supported by the European Union FP7 ITN INVISIBLES (Marie Curie Actions, PITN- GA-2011- 289442)and (HPRN-CT-200-00148); COST action MP1405 (Quantum Structure of Spacetime), COST action MP1210 (The String Theory Universe) as well as by FPA2012-31880 (MICINN, Spain)), FPA2011-24568 (MICINN, Spain), and S2009ESP-1473 (CA Madrid). The authors acknowledge the support of the Spanish MINECO {\em Centro de Excelencia Severo Ochoa} Programme under grant SEV-2012-0249.


\begin{thebibliography}{99}
\bibitem{note}
   
   	We restrict ourselves to dimension n=4 in what follows, in order not to clutter the formulas with n-dimensional Levi-Civita tensors. The results remain however valid in arbitrary dimension.
\end{thebibliography}
\end{document}